
\magnification=1100
\baselineskip 0.5truecm
\vsize 23.0 truecm
\hsize 14.5 truecm
\hoffset 0.5 truecm
\def\skuno{\vskip 20pt}
\def\skdue{\vskip 50pt}
\def\ltsim{\raise 2pt \hbox {$<$} \kern-1.1em \lower 4pt \hbox {$\sim$}}
\def\gtsim{\raise 2pt \hbox {$>$} \kern-1.1em \lower 4pt \hbox {$\sim$}}
\tolerance 3000

\parindent 0pt
\centerline {\bf VLBI Observations of a Complete Sample of Radio Galaxies}
\goodbreak
\centerline {\bf IV-The Radio Galaxies NGC2484, 3C109 and 3C382.}
\vskip 1.0truecm
\parindent 0pt
G. Giovannini$^{1,2}_7$, L. Feretti$^{1,2}_7$, T.Venturi$^2_7$, L. Lara$^3_8$,
J. Marcaide$^4_9$, M. Rioja $^3_{10}$,
\goodbreak
\parindent 0pt
S. R. Spangler$^5_{11}$, A. E. Wehrle$^6_{12}$
\skuno
\parindent 0pt
1) Dipartimento di Astronomia, Via Zamboni 33, 40100 Bologna, ITALY
\goodbreak
\parindent 0pt
2) Istituto di Radioastronomia, via Gobetti 101, 40129 Bologna, ITALY
\goodbreak
\parindent 0pt
3) Instituto de Astrofisica de Andalucia, CSIC, Apdo. 3004, 18080 Granada,
\goodbreak
\parindent 0.5truecm
SPAIN
\goodbreak
\parindent 0pt
4) Departamento de Astronomia, Universitat de Valencia, 46100 Burjassot,
\goodbreak
\parindent 0.5truecm
SPAIN
\goodbreak
\parindent 0pt
5) Department of Physics and Astronomy, University of Iowa, Iowa City,
\goodbreak
\parindent 0.5truecm
IA 52242 USA
\goodbreak
\parindent 0pt
6) Infrared Processing and Analysis Center, Jet Propulsion Laboratory,
\goodbreak
\parindent 0.5truecm
California Institute of Technology, M/S 100-22, Pasadena, CA 91125, USA
\vskip 2.0truecm
Postal Addresses:
\skuno
\parindent 0pt
7: Istituto di Radioastronomia, via Gobetti 101, 40129 Bologna, ITALY
\goodbreak
\parindent 0pt
8: Instituto de Astrof\'{\i}sica de Andaluc\'{\i}a, CSIC, Apartado 3004, 18080
Granada,
\goodbreak
\parindent 0.5truecm
SPAIN
\goodbreak
\parindent 0pt
9: Departamento de Astronomia, Universitat de Valencia, 46100 Burjassot,
\goodbreak
\parindent 0.5truecm
SPAIN
\goodbreak
\parindent 0pt
10: JIVE, c/o Radiosterrenwacht, Postbus 2, 7990 AA Dwingeloo
\goodbreak
\parindent 0.5truecm
The Netherlands
\goodbreak
\parindent 0pt
11: Department of Physics and Astronomy, University of Iowa, Iowa City
\goodbreak
\parindent 0.5truecm
IA 52242, USA
\goodbreak
\parindent 0pt
12: Infrared Processing and Analysis Center, Jet Propulsion Laboratory,
\goodbreak
\parindent 0.5truecm
California Institute of Technology, M/S 100-22, Pasadena, CA 91125, USA
\vfill
\eject
{\bf Summary}
\medskip
\parindent 20pt
We present here new VLBI observations of one  FR-I radio galaxy (NGC2484) and
two Broad Line FR-II radio galaxies (3C109 and 3C382). For 3C109 new VLA maps
are also shown. These sources belong to a
complete sample of radio galaxies under study for a better knowledge of their
structures at parsec resolution. The parsec structure of these
3 objects is very similar: asymmetric emission which we interpret as the
core plus a one-sided jet. The parsec scale jet is always on the
same side of the main kpc-scale jet. The limit on the jet to counterjet
brightness ratio, the ratio of the core radio power to the total radio
power and the Synchrotron-self Compton model allow us to derive
some constraints on the jet velocity and
orientation with respect to the line of sight.
{}From these data and from those published on 2 other sources of our sample,
we suggest that parsec scale jets are relativistic in both FR-I and FR-II radio
galaxies and that pc scale properties in FR-I and FR-II radio galaxies are
very similar despite of the large difference between these two  classes of
radio
galaxies on the kpc scale.
\vskip 1.5truecm
\parindent 0pt
{\bf Subject Headings:} Radio Continuum: Galaxies; Galaxies: Nuclei; Galaxies:
Jets;
Galaxies: Individual by Name (NGC2484, 3C109, 3C382, NGC315, 3C338);
techniques: interferometric
\vfill
\eject
{\bf 1. Introduction}
\medskip
\parindent 20pt
The knowledge of the structure of radio galaxies on the parsec scale is
relevant to test current models on jet dynamics and obtain new
information to confirm the proposed radio source "unified schemes".
Unification models suggest that FR-I radio galaxies
(see Fanaroff and Riley 1974 for the definition of FR-I and FR-II radio
galaxies) are the parent population
of X-ray and radio selected BL Lac objects. The distinction between
the three classes may correspond to different angles of the beamed emission
with respect to the line of sight.
Similarly, FR-II radio galaxies may be the
parent population of steep spectrum, lobe-dominated radio quasars and flat
spectrum, core-dominated quasars (Orr and Browne, 1982; Barthel, 1989;
Ghisellini et al., 1993; Antonucci, 1993).
\goodbreak
\parindent 20pt
To test these models and to get new insight on the knowledge of radio galaxies
on the parsec scale, we undertook a project of VLBI observations of a complete
sample of radio galaxies, selecting from the B2 and 3CR galaxy
samples those objects with a core flux density greater than 100 mJy at
6 cm at arcsecond resolution. This sample was presented in Giovannini et al.
(1990 - Paper I) and consists of 27 radio galaxies: 15 belong to the
FR-I class, 5 to the FR-II class and 7 are
unresolved or slightly resolved at the arcsecond resolution.
The core flux limit, imposed by observational constraints,
could in principle produce a sample biased towards objects with the core flux
density
enhanced by Doppler boosting. Therefore we could have a larger number of
objects
with jets pointing towards the observer. This point is not important in
discussing the single objects but will be taken in account and
better discussed in the future when we will consider the statistical
properties of the whole sample.
\goodbreak
\parindent 20pt
VLBI mapping and analysis of the radio galaxies of the sample
is in progress. A detailed study on the radio galaxy NGC315 (Venturi et al,
1993; paper II) and 3C338 (Feretti et al., 1993; paper III), have been
published. Here we present new VLBI observations on 3 more radio galaxies
of our sample: NGC2484, 3C109 and 3C382.
\goodbreak
\parindent 20pt
In this paper we use a Hubble constant H$_0$ = 100 km sec$^{-1}$ Mpc$^{-1}$
and
\goodbreak
\parindent 0pt
q$_0$ = 1.
\vskip 1.0 truecm
{\bf 2. The Sources}
\vskip 1.0 truecm
{\it 2.1 NGC2484}
\medskip
\parindent 20pt
The radio galaxy ~~NGC2484 (B2 0755+37) is a bright E galaxy ~~~(M$_v$ = -21.9)
with a redshift of 0.0413 (Colla et al. 1975), which gives the following
linear size conversion: 1 milliarcsecond (mas) corresponds
to 0.57 parsec (pc). At
arcsecond resolution, it shows
a flat spectrum core with two extended lobes in position angle (PA) 110$^\circ$
and it is classified as a FR-I radio
source. Its total radio power at 408 MHz is 1.10$\times$10$^{25}$ W/Hz while
the radio power of the arcsecond core at 5.0 GHz is 3.55$\times$10$^{23}$ W/Hz.
An obvious jet is imbedded in the SE lobe. At higher angular resolution,
a very short jet is also visible in the North-West (NW) direction. It is well
collimated
at the beginning, but it widens at a short distance from the nucleus to
form the radio lobe (de Ruiter et al. 1986; Parma et al. 1987).
The radio emission is highly polarized at 1.4 GHz (up to 30\%) and
the bright jet points to the more polarized lobe (Capetti et al. 1993)
suggesting that the main jet is oriented towards the observer and that its
luminosity is enhanced
by Doppler effects (Parma et al., 1993).
The arcsecond core flux density was monitored for 5 years (Ekers et al., 1983),
but no variability was found.
\goodbreak
\parindent 20pt
For this galaxy we present new VLBI data at 6 cm.
\vskip 1.0 truecm
{\it 2.2 3C109}
\medskip
\parindent 20pt
The radio source 3C109 has been identified with an isolated, Broad Line
N galaxy (Grandi and Osterbrock, 1978), with a redshift of 0.3066 (Goodrich and
Cohen, 1992) and M$_v$ = -22.9. The linear size conversion is 2.6 pc/mas. A
recent optical observations (Goodrich and Cohen, 1992) have suggested that
its nuclear emission is obscured by dust. According to these authors,
if the nucleus were seen unobscured, the surrounding
galaxy would not be visible because of the very strong nuclear light and
3C109 would be classified as an intrinsically bright quasar. It is dimmed
along the
line of sight by intervening dust and the relativistic beamed structure
should not lie close to the plane of the sky.
\goodbreak
\parindent 20pt
The presence of high internal absorption in 3C109 is confirmed
by recent ROSAT X-ray observations presented by Allen and Fabian (1992).
In fact, they find an excess column density associated with the AGN of
$\sim$ 5$\times$10$^{21}$ atom cm$^{-2}$ (redshift-corrected).
This value is consistent with the reddening observed in optical studies.
Therefore the interpretation of the nature of 3C109 is shared by
Allen and Fabian (1992), who conclude that 3C109 is identical to
luminous steep spectrum quasars and it is classified as a galaxy due to its
intrinsic absorption.
They suggest an intermediate angle of $\sim$ 50$^\circ$ between
the radio axis of 3C109
and the line of sight assuming that the X-ray absorption is due to a torus of
obscuring material surrounding the AGN with its symmetry axis along the radio
axis.
\goodbreak
\parindent 20pt
The large scale radio structure consists of two
very symmetric lobes with hot spots and core radio emission. It is a typical
FR-II radio galaxy; the total radio power at 408 MHz is 1.32$\times$10$^{27}$
W/Hz (it has the highest radio power in our sample) and the arcsecond core
radio
power at 5.0 GHz is 1.32$\times$10$^{25}$ W/Hz (Giovannini et al., 1988).
The flux density ratio between the two lobes is 1.0 $\pm$ 0.1
and the distance ratio between the core and the two hot spots is
1.00 $\pm$ 0.03.
\goodbreak
\parindent 20pt
The arcsecond radio core showed variability in the range
180 - 400 mJy (Ekers et al., 1983) in the period 1974 - 1980.
Simultaneous observations with the Very Large Array (VLA)
by Antonucci (1988) at 6 and 1.3 cm, give a flux
density at 6 cm
of 234.5 mJy ~~~with a spectral ~~~index $\alpha^{1.3}_6$ = 0.26
{}~~~(S($\nu$) $\propto$ $\nu$ $^{-\alpha}$).
Evidence for variability at optical and near-infrared wavelengths has been
reported by Rudy et al. (1984); Allen and Fabian (1992) report long term
X-ray variability.
\goodbreak
\parindent 20pt
We observed 3C109 with VLBI at 6 cm to investigate the jet morphology
in the inner parsec region.
We also present VLA observations obtained to compare the parsec and kiloparsec
scale radio morphologies. These VLA observations
also reveal the polarization structure of the two lobes.
\vskip 1.0 truecm
{\it 2.3 3C382}
\medskip
\parindent 20pt
The radio source 3C382 (B2 1833+32) is identified with a D3 type galaxy
(Matthews et al., 1964). Optical spectrophotometry by Osterbrock et al.
(1976) shows that the nucleus has a high-excitation forbidden-line
spectrum
with redshift z=0.0578 superimposed on broad permitted lines and a strong
continuum, typical of a Broad Line Radio Galaxy. It is peculiar in possessing
very broad (FWHM $\sim$ 30000 km s$^{-1}$) emission lines (Tadhunter et al.,
1986). Its optical magnitude is M$_v$ = -22.2. For this galaxy the linear size
conversion is 0.76 pc/mas.
\goodbreak
\parindent 20pt
Exosat X-ray observations showed a doubling time of the X-ray flux of 3 days,
consistent with the general correlation of X-ray luminosity with variability
time scale for Seyferts and QSO's (Kaastra et al., 1991 and references
therein).
\goodbreak
\parindent 20pt
On the arcsecond scale its radio structure consists of a
compact, flat spectrum core coincident with the galaxy and two radio lobes
aligned in PA $\sim$55$^\circ$ (Strom et al. 1978, Leahy et al. 1991,
Black et al., 1992).
The total radio power at 408 MHz is 5.01$\times$10$^{25}$ W/Hz and it shows
the classical morphology of FR-II radio galaxies.
In the north-eastern lobe a faint jet is visible on the gray-scale map given by
Black et al. (1992). The jet is not straight from the core to the hot spot, but
shows a gradual bend. On the other side of the core no jet-like structure
has been detected.
\goodbreak
\parindent 20pt
3C382 is known to be variable at X-ray (Barr and Giommi 1992 and
references therein), radio (Strom et al. 1978), optical (Puschell 1981)
and ultraviolet
(Tadhunter et al. 1986) wavelengths. In the radio regime
Strom et al. (1978) report a flux density decrease from 230 mJy to 175 mJy in
the period 1972.5 - 1974.3; Ekers et al. (1983) however, found a constant core
flux density of 206 $\pm$12 mJy in the period from 1976.0 to 1980.2.
\goodbreak
\parindent 20pt
We present here new VLBI observations at 2.3 and 8.4 GHz of the
nuclear
region of this galaxy. At 8.4 GHz a high resolution map is available
while at 2.3 GHz only a model fit to the visibility data is possible due to the
lack of data in many baselines (see Sect. 3.1).
\vskip 1.0truecm
{\bf 3. Observations and Data Reduction}
\medskip
{\it 3.1 VLBI data}
\medskip
\parindent 20pt
The observational details for each source are given in Table 1:
source name (col. 1), observing frequency
(col. 2), observing mode and bandwidth (col.3 and 4 respectively);
telescopes and observing dates (col. 5 and 6 respectively).
\goodbreak
\parindent 20pt
For the sources NGC2484 and 3C109, after the data correlation and fringe
search, the output visibilities were written in FITS format.
Initially the visibility amplitudes were calibrated using the gain values and
system temperatures provided by the antenna during the observations.
Later the gain values were checked, and when necessary
corrected using standard VLBI calibrator sources assumed to be unresolved
by the European and US baselines. The calibrator flux density was
measured during the observations at the Effelsberg and Owens Valley telescope.
The corrections to the ~~~~given antenna gains were
{}~~~~always ~~~~\ltsim 10\%. The calibrated data were global
fringe fitted in AIPS and reduced using the AIPS package with the standard
procedure:
as a first step the data were edited to
delete the bad points. A map was preliminary obtained whose clean components
were used as the input model for the phase self-calibration cycle.
When good agreement between the data and the source model was reached and
phases were stable, we
allowed gain self-calibration with 2 hours time scale to correct amplitude
calibration errors.
\medskip
\parindent 20pt
The radio galaxy 3C382 was observed at 2.3 and 8.4 GHz simultaneously.
The experiment consisted of observations of the quasar 3C395  phase-referenced
to the core of 3C382. The switching duty cycle between the two sources only
allowed for scans shorter than 5 minutes on 3C382 which, however, were well
spread in the u-v plane.
At Effelsberg and the VLA only the 8.4 GHz receiver was available
(see Table 1).
After the global fringe fitting the data
were exported to Caltech package format for mapping purposes
(Pearson, 1991) and to a format compatible with the astrometric program VLBI3
(Robertson, 1975). The astrometric part of the work will be reported elsewhere
(Lara et al., in preparation).
Standard calibration, editing and hybrid mapping procedure were used to obtain
our final image.
\goodbreak
\parindent 20pt
At 2.3 GHz 3C382 was detected only
on the two baselines Noto-Medicina and Medicina-Onsala due to technical
problems and the low correlated flux density. Therefore, only
model fitting to the data on these baselines was possible
at this frequency.
\vskip 1.0 truecm
{\it 3.2 VLA data}
\medskip
\parindent 20pt
The radio galaxy 3C109 was observed with the VLA in A and B configurations
at 1.4 GHz and in B and C configurations at 4.9 GHz in 1982 and 1983,
in the same observing sessions as reported
earlier for 3C411 (Spangler and Pogge, 1984) 3C79 and 3C430
(Spangler et al., 1984). Details of the VLA observations may be found in those
papers. Maps were obtained combining the A and B arrays at 1.4 GHz and the
B and C arrays at 4.9 GHz to have the same UV-coverage at the 2 frequencies.
We obtained also a 6 cm map in the A configuration using the data
recorded at the VLA during VLBI observations. The data were calibrated
in the standard way using the AIPS package. We observed 3C48 as gain calibrator
just before the start of the VLBI session. The calibrator
source 0406+121 was observed every 30 minutes to phase the array. The final map
was
obtained synthesizing a large field (2048$\times$2048 pixels) due to the large
size of
3C109. Data were first edited and then self-calibrated for phase and
amplitude.
In the final map, the noise level is 0.05 mJy/beam with a resolution of 0.55
arcsecond (HPBW).
For technical reasons, no polarized data are available at this resolution.
An attempt to combine these data with previous 4.9 GHz data, to obtain a high
resolution map with a good uv-coverage also at the short baselines
(A+B+C
configurations), was not possible due to problems in homogenizing the data
and to the variability of the core flux density.
\vskip 1.0truecm
{\bf 4. Results}
\medskip
\parindent 20pt
Figure 1 shows the final uv coverage for all the 3 sources, except for the data
on 3C382 at 2.3 GHz. In Table 2 some VLBI observational parameters and results
are summarized for each source; columns 1 - 9 contain:
the source name and the observing frequency (col. 1 and 2), the synthesized
HPBW, the major axis PA and the noise level (col 3, 4 and 5), the total VLBI
flux density and the core flux density (col 6 and 7) and the jet flux density
and PA (col 8 and 9 respectively).
\medskip
{\it 4.1 NGC2484}
\medskip
\parindent 20pt
A contour plot of the VLBI map is shown in Fig. 2, at full resolution (2a) and
with a slightly super-resolved and round beam (2b).
At parsec resolution
NGC2484 shows an elongated radio structure. No spectral index information is
available, however on the basis of the flat
spectrum of the arcsecond core ($\alpha_{5.0}^{1.4}$ = 0.07), we can reasonably
assume the parsec scale core
as coincident with the brightness peak in the VLBI map. In this case, the
morphology
is asymmetric with an extended feature oriented in PA 118$^\circ$
$\pm 4^\circ$. This feature is not in agreement with the jet definition given
by Bridle and Perley (1984).
However, due to the source structure similarity with other parsec scale
structures and being the small jet
oriented as the main kpc scale jet, we interpret it as a one-sided jet.
The jet-counterjet brightness ratio is \gtsim 20 at 2 mas from the core.
The total flux in the VLBI map is 216 mJy (see Table 2), in good agreement with
the correlated flux density in our
shortest baseline. The core flux density at the arcsecond resolution is 195 mJy
(see Giovannini et al., 1988).
The discrepancy is within the uncertainties present in the VLBI flux scale
calibration (10\%).
We have deconvolved the VLBI structure with 2 Gaussians to derive the
core flux density. We estimate that the parsec scale core has a flux of
$\sim$140 mJy which matches with the correlated flux density in the longest
baselines (e.g. Effelsberg-Owens Valley). The jet PA in the VLA map is
110$^\circ\pm$2$^\circ$ (marginally different from
the pc scale jet PA (118$^\circ$ $\pm 4^\circ$)).
\goodbreak
\parindent 20pt
\medskip
{\it 4.2 3C109}
\medskip
{\it 4.2.1 Large Scale Structure}
\medskip
\parindent 20pt
In Figures 3 and 4 the 1.5 arcsecond maps obtained at the VLA at 20 and 6 cm,
respectively, are shown. On the 6 cm map the polarization vectors are
superimposed. In Figure 5 the map obtained at the VLA with a resolution
of 0.55 arcsecond, using the A configuration is shown.
Note that the maps in Figures. 3 and 4 have been rotated by 55$^{\circ}$
for display purposes.
\goodbreak
\parindent 0pt
The projected linear size of 3C109 from our maps is $\sim$ 280 kpc. The hot
spots are clearly visible and substructures are present in the 2 lobes.
The distance of the two hot spots from the core is the same.
A faint jet is present in the South-Eastern (SE) direction connecting the
core to the SE lobe.
The jet is well defined near the core and within the SE lobe but is only
marginally detected in the region between the core and the hot spot.
The jet is not straight, but near the core it oscillates ($\pm$10$^\circ$)
around the PA 150$^\circ$ while in the lobe it starts to bend to reach
a final PA (near the hot spot) of $\sim$90$^\circ$. No jet-like feature is
visible on the other side of the source core. The jet-counterjet brightness
ratio is \gtsim 24 at 1 arcsecond from the core.
The low resolution maps confirm the symmetry in size found in the high
resolution map, and show an extended low brightness region
between the core and the hot spots.
\goodbreak
\parindent 20pt
The radio structure of the two hot spots is very interesting (Fig. 5).
The SE one is the most compact in agreement with Laing (1989) who noted that
compact hot spots are seen only on the jetted side of FR-II radio sources;
two secondary knots, which mark the path of the
jet, are present at $\sim$4 and
10 arcsecond of distance from the main hot spot. The NW hot spot is elongated
perpendicularly to the hot spot-core direction; a 'S' shaped structure is
visible
inside the inner region of the lobe.
\goodbreak
\parindent 20pt
We used the maps obtained combining the A and B arrays at 1.4 GHz and
combining the B and C arrays at 4.9 GHz to derive the
polarized flux from 3C109. In Table 3 we present
the total flux density and the polarization percentage, derived from the maps
with HPBW = 1.5 arcsecond at 6 and 20 cm; values are given for:
the whole southern (col. 1) and northern (col. 5) lobe, the hot spot regions
(defined as the regions where radio emission is visible in the high resolution
map shown in Fig. 5; col. 2 and 6), the diffuse regions (col. 3 and 7) and the
core (col. 4).
The two hot spot regions are polarized at a level
of 10\% while no polarized flux is visible within the errors in the diffuse
regions and in the core source.
The polarization percentage is the
same in the two lobes at the two frequencies; also the PA of the polarized
vectors does not change.
This result is only apparently in contradiction with the Laing-Garrington
effect
(Laing 1988; Garrington et al., 1988)
and the suggestion (see Sect. 2.2) that 3C109 is not on the plane of the sky.
In fact Garrington and Conway (1991) estimated for a typical
depolarizing halo a core radius of $\sim$ 100 kpc for low redshift (z $<$ 1)
sources. Since we detect
a polarized flux only in the external regions of this double source,
given the large projected size of 3C109 (280 kpc), we can reasonably assume
that the polarized radio lobes are outside the more dense regions of the galaxy
corona of hot gas responsible of the depolarization.
Therefore we do
not see the Laing-Garrington effect in 3C109 simply because the polarized
flux comes from regions external to the halo surrounding 3C109 (if any).
This is consistent with the anticorrelation found by Garrington and Conway
(1991) between the depolarization asymmetry and the source linear size.
\medskip
{\it 4.2.2 Parsec Scale}
\medskip
The contour plot of the 4.9 GHz VLBI image at full resolution is shown
in Fig. 6a. Parameters of the radio source are given in
Table 2.
At VLBI resolution 3C109 shows a one-sided morphology, with
a bright core extended in
PA 130 $^\circ\pm$ 5$^\circ$, and a faint knot of radio emission aligned
in the same direction.
This feature is more obvious in a super-resolved map, convolved
with a round beam (Fig. 6b). Although this knot
does not match all the criteria necessary to classify it as a jet (Bridle and
Perley, 1984), it is certainly related with the existence of a jet, then
we consider plausible to assume that on the parsec scale
3C109 has a one-sided core-jet structure. The jet is faint and slightly
misaligned with respect to
the kpc scale jet. The jet/counterjet brightness ratio is \gtsim 7 at
$\sim$2 mas from the core.
\goodbreak
\parindent 20pt
The total flux density in the parsec
scale structure is in good agreement with the core flux density measured in the
VLA map obtained in the same epoch.
We point out that our VLBI observations were made when the core flux density
was at a minimum ($\sim$ 180 mJy).
\medskip
{\it 4.3 3C382}
\medskip
\parindent 20pt
In Figure 7 we present the 8.4 GHz VLBI image of 3C382.
The total flux density of the VLBI
structure is 190 mJy with a compact core of 115 mJy.
\goodbreak
\parindent 20pt
The VLBI structure of 3C382 presents a core-jet morphology with marked
wiggles:  the jet extends in PA $\sim$40$^\circ$ for $\sim$ 1 mas,
then in PA $\sim$90$^\circ$ up to $\sim$3 mas from the core and finally again
in PA $\sim$40$^\circ$.
In Table 2 we give the map parameters.
The jet/counterjet brightness ratio is \gtsim 30 at $\sim$2 mas from the core.
\goodbreak
\parindent 20pt
The parsec scale jet is on the same side of the faint kpc-scale jet visible on
the map given by Black et al. (1992). The kpc-scale jet PA is $\sim$
50$^\circ$, but, as in 3C109, an evident change in the PA is necessary to
reach the hot spot (the final PA is 90$^\circ$), confirming an instability
in the jet position angle from the pc to the kpc scale in this source.
\goodbreak
\parindent 20pt
At 2.3 GHz the best model obtained from modelfitting the two baselines
Medicina-Noto and Medicina-Onsala
(see Sect. 3.1) gives a  single Gaussian component of 157 mJy flux density
and slightly extended in
PA $\sim$50$^\circ$ consistent
with the results at 8.4 GHz. An estimate of the overall spectral index
between the two frequencies gives $\alpha$ $\sim$ -0.1.
\vskip 1.0truecm
{\bf 5. Discussion}
\medskip
\parindent 20pt
The three radio galaxies presented here show similar parsec scale morphologies:
a strong compact component, identified with the core, and a faint
one-sided jet. In all cases the parsec scale jet is oriented on the same side
of the kpc scale main jet. This correlation implies either that jets are
intrinsically asymmetric or that parsec and kpc scale jets are
relativistic.
\goodbreak
\parindent 20pt
The presence of relativistic jets in strong radio sources as
quasars and FR-II radio galaxies
is now widely accepted (see Antonucci 1993 for a recent review).
Also, evidence is growing that radio jets in FR-I galaxies are initially
relativistic. This is consistent with the recent result by Parma et
al. (1993a) that in FR-I radio galaxies the stronger jet
is generally on the same side of the less-depolarized lobe, just as in the
FR-II
radio sources. This is also consistent with the decrease of jet sidedness
ratio with distance
from the core and its trend with the total and core radio power examined by
Parma et al. (1993b).
Laing (1993) developed a two component model in which
FR-I jets are relativistic on small (parsec)
scales, but decelerate quickly and become non-relativistic on large (kpc)
scales. This model explains the correlation between polarization and
sidedness in FR-I jets.
\goodbreak
\parindent 20pt
Therefore we interpret the radio
structures presented here as affected by
Doppler favoritism and will use the available data to constrain the possible
values of the intrinsic jet velocity and of the orientation of the radio source
with respect to the line of sight.
\goodbreak
\parindent 20pt
We also include in the discussion the data on two previously studied
sources belonging to our sample, i.e. 3C338
(Feretti et al., 1993) and NGC315 (Venturi et al., 1993).
In Venturi et al. (1993) we took into account the possibility that the
one-sided small scale morphology of NGC315 is caused by intrinsic asymmetries.
As stated before, we
adopt here the view that the jet is Doppler boosted.
This is also suggested by Bicknell (1994) who modeled the
radio jets of NGC315 assuming an initially relativistic velocity and
deceleration by turbulent entrainment.
\goodbreak
\parindent 20pt
Including 3C338 and NGC315, our results concern two FR-II and three
FR-I radio galaxies.
\medskip
{\it 5.1 Jet Sidedness}
\medskip
\parindent 20pt
Assuming that the jets are intrinsically symmetric, from the jet to counterjet
brightness ratio R we can constrain
the value of $\beta cos\theta$, according to the formula
\goodbreak
\centerline {R = (1 + $\beta cos\theta$)$^{2+\alpha}$
(1 - $\beta cos\theta$)$^{-(2+\alpha)}$}
\goodbreak
\parindent 0pt
where $\beta$ is the ratio of the jet velocity to the speed of light and
$\theta$ is the angle to the line of sight. The jet spectral
index $\alpha$ is assumed to be 0.5 (Pearson and Zensus, 1987). The validity of
this standard
formula depends on the degree of isotropy of the intrinsic synchrotron
emissivity in the
jet (see Begelman, 1993). We may expect some anisotropy in the
observed radiation depending on the degree of ordering and the magnetic field
orientation.
The measure of magnetic field properties may be derived from maps of the
polarized flux at parsec resolution. In a jet with
20\% polarized flux, the anisotropy due to the magnetic field will increase
the beaming ratio by less than 30-50 \%. The effect can be much more prominent
in highly polarized jets ($\sim$ 50\%) with an increase of a factor 2-3 in
the beaming ratio
(Begelman, 1993). In the following discussion we shall assume that the
fractional
polarization in the parsec scale jet is not too large, i.e.
that the parsec scale jet emissivity
is roughly isotropic. This assumption has to be tested in the future with VLBI
maps of the polarized flux in a large sample of radio galaxies.
\goodbreak
\parindent 20pt
We have obtained the value of $\beta cos\theta$, from the VLBI jet and
counterjet ratios. In the sources 3C109
and NGC2484 some brightness asymmetry is present also on the arcsecond scale.
By assuming
that this asymmetry still originates from Doppler boosting, the factor
$\beta$cos$\theta$ can be constrained also on this scale. In particular the
constraint on the
arcsecond scale for 3C109 is stronger than that on the parsec scale
owing to the better signal to noise ratio of the VLA map.
Since we do not consider a jet
reacceleration from the parsec to the kpc scale, for this source
we will adopt the limit on $\beta$cos$\theta$ derived from the VLA map
for both scales.
For NGC2484 the arcsecond
constraint is in agreement with a jet deceleration from the parsec to the kpc
scale. No discussion is possible on the kpc scale jet of 3C382 for the lack of
a jet counterjet ratio on this scale.
\goodbreak
\parindent 20pt
The obtained values of $\beta cos\theta$ are
summarized in Table 4 which is organized as follows: col. 1 the
source name, cols. 2, 3 and 4 respectively the VLBI
jet to counterjet brightness ratio R,
the parsec-scale limit to $\beta cos\theta$ and the distance 'd' in
milliarcsecond between
the core and the region where R was estimated; cols. 5,6 and 7 the same
values from the VLA maps with the distance in arcsecond.
In Fig. 8 the allowed region in the $\theta - \beta$ space is shown.
\medskip
{\it 5.2 Core Radio Power}
\medskip
\parindent 20pt
Given the existence of a general correlation between the core and the total
radio power in radio galaxies (Giovannini et al., 1988), we can derive an
expected intrinsic core radio power from the total galaxy radio power.
Since the total radio power is measured at low frequency and  is
therefore not affected by Doppler boosting, the core emission implied by the
total radio power is not boosted. Objects with
a core radio power stronger than the expected value could be interpreted as
galaxies where the core radio emission is contaminated by a Doppler boosted
relativistic jet lying close to the line of sight. Therefore a constraint on
$\beta cos\theta$ can be derived independently from the jet sidedness.
Giovannini et al. (1988) found the following correlation between the
arcsecond core
radio power P$_c$ and the total radio power P$_t$ using a sample of 187 radio
galaxies selected at low frequency:
\goodbreak
\centerline {Log P$_c$ = 11.01($\pm$1.05) + 0.47($\mp$0.4)LogP$_{tot} $}
\goodbreak
\parindent 0pt
We expect that the radio galaxies used to derive the previous correlation are
at different angles with respect to the line of sight and have an uniform
distribution being selected at low frequency. In this case the dispersion of
P$_c$ around the best fit value reflects the different orientation to the
line of sight. Assuming that sources are oriented at random angles, the best
fit value corresponds to the average orientation angle
(60$^\circ$) with respect to the line of sight and the previous
correlation should be read as:
\goodbreak
\centerline {Log P$_c$(60) = 11.01 + 0.47LogP$_{tot} $}
\goodbreak
\parindent 0pt
where P$_c$(60) is the apparent (beamed) core radio power for a galaxy oriented
at 60$^{\circ}$ with respect to the line of sight.
Since the Doppler enhancement in a jet oriented at an angle $\theta$ is:
\goodbreak
\centerline {P$_c$($\theta$) = P$_i$ (1 - $\beta cos\theta$)$^{-(2 + \alpha)}$}
\goodbreak
\parindent 0pt
where P$_i$ is the intrinsic radio power and P$_c$($\theta$) is the apparent
(beamed) radio power,
assuming $\alpha$ = 0 for the core emission, we finally obtain:
\goodbreak
\parindent 0pt
\centerline {$\beta$ = (K - 1) (K cos$\theta$ - 0.5)$^{-1}$}
\goodbreak
\parindent 0pt
with K = [P$_c(\theta)$/P$_c$(60)]$^{0.5}$.
We used this correlation to derive upper and lower limits to $\beta$ and
$\theta$, taking into account the statistical uncertainties (1 $\sigma$) and a
possible core flux density variability up to a factor of 2.
The derived values are drawn in Fig. 8.
\goodbreak
\parindent 20pt
We point out that in the above argument the core flux density measured
on the arcsecond scale is used.
A better estimate of
the beaming factor of the core could be obtained using the VLBI core flux
density, but the main problem to do this is that the correlation
between the total radio power and the milliarcsecond core radio power is poorly
known and still affected by large uncertainties. In our Paper I
(Giovannini et al., 1990) we derived this correlation for 20 objects
and we found it to be consistent within the errors with the one used in the
present discussion.
\medskip
{\it 5.3 The Synchrotron-self Compton emission}
\medskip
\parindent 20pt
Independent constraints on the amount of beaming for a radio source can be
derived from the
Synchrotron-self Compton (SSC) model of X-Ray emission from
the nuclear region (see Marscher, 1987; Ghisellini et al., 1993).
In principle, when the core angular size is known, the comparison between
the predicted and observed
X-Ray flux density gives a lower limit to the Doppler factor $\delta$ defined
as:
\goodbreak
\centerline {$\delta$ = [$\gamma(1 - \beta cos\theta)]^{-1}$}
\goodbreak
\parindent 0pt
where $\gamma$ = (1 - $\beta^2$)$^{-1/2}$.
\goodbreak
\parindent 20pt
We derived $\delta$ using
the Ghisellini et al. (1993) formula assuming that the core emission is
due to a partially opaque electron synchrotron radiation from a uniform
sphere and that $\alpha$ = 0.5 for the thin synchrotron emission.
The parameters entering in the computation are given in Table 5 as follows:
redshift, VLBI core
size D$_{VLBI}$, radio flux F$_m$ at the self absorption frequency $\nu_m$ and
X-Ray flux F$_x$ at the energy h$\nu_x$.
We want to note that the derived values of $\delta$, given in Table 5, are
crude estimate due to the uncertainties in the knowledge of
the self absorption radio frequency, the core angular size and
the X-Ray flux. In particular, the X-Ray flux can be overestimated due to the
contamination of a thermal component which is found to be present in radio
galaxies (see e.g. Worral and Birkinshaw, 1994). For this reason we use the
Einstein HRI flux when available (only for NGC315).
\goodbreak
\parindent 20pt
Lower limits of $\delta$ obtained here, are in agreement with
those given in Ghisellini
et al. (1993) for radio galaxies and lobe-dominated quasars confirming that
radio galaxies have much smaller $\delta$ than compact quasars.
Only for NGC315 the lower limit of $\delta $ is $>$ 1 allowing us to
constrain $\beta$ and $\theta$ (see Fig. 8).
For the other sources the derived lower limits are not conclusive.
\medskip
{\it 5.4 Implications on the jet velocity and orientation}
\medskip
\parindent 20pt
The implications  on the jet velocity and orientation come from the constraints
obtained in Sect. 5.1, 5.2 and 5.3.
For each source we also imposed an upper limit on the angle to the line of
sight $\theta$, to restrict the maximum intrinsic radio source size to
1.5 Mpc.
\goodbreak
\parindent 20pt
As discussed before, a larger range of allowed values is possible
from the jet/counterjet brightness ratio if we
assume that some anisotropy, due to the directionality of
the magnetic field, is present in the intrinsic synchrotron jet
emissivity (Begelman, 1993). However, we note that in our sources the major
constraint to the values of
$\beta$ and $\theta$ is obtained from the dominance of the radio
core on the extended radio emission.
\goodbreak
\parindent 20pt
While it seems difficult to
further improve the constraints on $\beta$ and $\theta$
from the ratio P$_c(\theta)$/P$_c$(60), due to the flux density variability
of the radio galaxy cores, we expect to derive stronger constraints
in the future from the ratio between the jet and counterjet brightness,
which is presently limited by the noise level in the maps.
Thanks to the better sensitivity reachable now with the global array
(European VLBI Network  + Very Long Baseline Array),
we should obtain maps with a noise level low enough to put stronger
constraints on $\beta$ and $\theta$.
Moreover high resolution X-Ray data from the Rosat satellite could be helpful
for the application of the SSC model.
\goodbreak
\parindent 20pt
{}From a careful analysis of figure 8, we can summarise our results as follows:
\goodbreak
\parindent 0pt
- the allowed range of $\theta$ is 10$^\circ$ $<$ $\theta$ $<$ 35$^\circ$ for
3C109.
This result confirms even at radio frequencies that 3C109 is an obscured quasar
as suggested by X-Ray and optically observations (see Sect. 2.2).
This is in agreement with
the prediction of unified models where lobe dominated broad line radio galaxies
are interpreted as lobe dominated QSS which appear as galaxies
only because nearby (see e.g. Antonucci, 1993) and therefore are
expected to be at 10$^\circ$ $<$ $\theta$ $<$ 40$^{\circ}$ with respect to
the line of
sight (see e.g. Ghisellini et al., 1993). In the case of 3C382, which
is a lobe dominated broad line radio galaxy as 3C109, our constraints
are not as strong as for 3C109 and the allowed range of $\theta$ is
5$^\circ$ $<$ $\theta$ $<$ 48$^\circ$;
\goodbreak
\parindent 0pt
- for NGC315 the allowed range of $\theta$ is narrow
(30$^\circ$ $<$ $\theta$ $<$ 41$^\circ$). This range is fully consistent with
the unification schemes, which predict FR-I radio galaxies to be at angles
$\theta$ \gtsim ~30$^{\circ}$ to the line of sight (see e.g. Ghisellini et
al., 1993). For NGC2484 instead the present data cannot exclude angles
$<$ 30$^{\circ}$;
\goodbreak
\parindent 0pt
- the case of 3C338 is rather peculiar due to the symmetry of the radio
structure: all angles are allowed even if small orientation angles imply low
jet velocities;
\goodbreak
\parindent 0pt
- the possible range of $\beta$ is 0.8 \ltsim ~$\beta$~ \ltsim 1 for NGC315 and
3C109 and 0.6 \ltsim ~$\beta$~ \ltsim 1 for NGC2484 and 3C382. For
3C338 we derive lower values of $\beta$
infact either the source lies in the plane of the sky and
$\beta$ \ltsim 0.4 or the jets are not relativistic and all angles are
allowed.
We note however that the $\beta$ upper limit for 3C338 derives from the ratio
between the core and total radio power and that the 3C338 core emission
is strongly variable (see Feretti
et al., 1993). Therefore the uncertainty from the core - total radio
power correlation may be larger than the one taken in account here.
\goodbreak
\parindent 20pt
No reliable map at a different
epoch is presently available to search for a possible
superluminal motion in these sources. Assuming a high jet velocity ($\beta$
\ltsim 1) we expect for these sources a proper motion \ltsim 1 mas/yr.
Observations have been already planned and will give soon useful data
to discuss this point.
\goodbreak
\parindent 20pt
We also note that
at parsec resolution we do not see any difference between FR-I and FR-II
radio galaxies. In both classes of sources, the observed radio
structures are very similar and may be explained assuming that parsec scale
jets are relativistic.
Data on more radiogalaxies are necessary to confirm this similarity
between FR-I and FR-II radio galaxies on the parsec scale and to study
possible correlations between the nuclear power, the jet velocity and
the large scale radio morphology. However, this similarity could indicate
that the large morphological difference between FR-I and FR-II radio galaxies
on the kpc scale is due to a different interaction of the jet with the
surrounding medium on a scale larger than the VLBI one.
\goodbreak
\parindent 20pt
We finally wish to comment on the flip-flop model, which was invoked as a
possible interpretation of the structure of 3C338 (Feretti et al., 1993).
No evidence in favour of
this model was found in other sources studied so far.
Therefore it seems that even in
the case that the jets are fed by an alternate ejection, the jet speed must
be relativistic to account for the one-sidness of the source structures
on the parsec scale.
\vfill
\eject
{\bf Acknowledgement}
\medskip
\parindent 20pt
We acknowledge R. Fanti for helpful discussions and suggestions, C. Fanti and
F. Mantovani
for a critical reading of the manuscript and an anonymous Referee for many
useful comments and suggestions
which improved this work. We thanks the staffs at the telescopes
for their assistance with the observations and the people assisting us in the
data correlation in Bonn and at the California Institute of Technology.
Thanks are due to Mr. V. Albertazzi for drawing the figures.
\goodbreak
\parindent 20pt
L.F. and G.G. acknowledge financial support from Italian MURST and
J.M, L.L. and M.R. acknowledge for a grant (PB89-0009) from Spanish CICYT.
\goodbreak
\parindent 20pt
The National Radio Astronomy Observatory is operated by Associated Universities
Inc. under contract with the National Science Foundation
\vfill
\eject
{\bf References}
\medskip
\parindent 0pt
Allen S. W., Fabian A. C.: 1992 MNRAS 258, 29p
\goodbreak
\parindent 0pt
Antonucci R. R. J.: 1988 ApJ 325, L21
\goodbreak
\parindent 0pt
Antonucci R. R. J.: 1993 ARA\&A 31, 473
\goodbreak
\parindent 0pt
Barr P., Giommi P.: 1992 MNRAS 255, 495
\goodbreak
\parindent 0pt
Barthel P. D.: 1989 ApJ 336, 606
\goodbreak
\parindent 0pt
Begelman M. C.: 1993 in Jets in Extragalactic Radio Sources, Roeser H.-J and
\goodbreak
\parindent 20pt
Meisenheimer ed., p 145
\goodbreak
\parindent 0pt
Bicknell, G. V., de Ruiter H. R., Fanti R., Morganti R., Parma P.: 1990
\goodbreak
\parindent 20pt
ApJ 354, 98
\goodbreak
\parindent 0pt
Bicknell, G. V.: 1994 ApJ in press
\goodbreak
\parindent 0pt
Bierman P.I., Kuehr H., Snyder W.A., Zensus J.A.: 1987 A\&A 185, 9
\goodbreak
\parindent 0pt
Black A. R. S., Baum S. A., Leahy J. P., Perley R. A., Scheuer P. A. G.: 1992
\goodbreak
\parindent 20pt
MNRAS 256, 186
\goodbreak
\parindent 0pt
Bridle A. H., Perley R. A.: 1984 ARA\&A 22, 319
\goodbreak
\parindent 0pt
Capetti A., Morganti R., Parma P., Fanti R.: 1993 A\&AS in press
\goodbreak
\parindent 0pt
Colla G., Fanti C., Fanti R., Gioia I., Lari C., Lequeux J., Lucas R., Ulrich
M.
\goodbreak
\parindent 20pt
H.: 1975 A\&A 38, 209
\goodbreak
\parindent 0pt
de Ruiter H. R., Parma P., Fanti C., Fanti R.: 1986 A\&AS 65, 111
\goodbreak
\parindent 0pt
Ekers R. D., Fanti R., Miley G. K.: 1983 A\&A 120, 297
\goodbreak
\parindent 0pt
Fabbiano G., Miller L., Trinchieri G., Longair M., Elvis M.: 1984 ApJ 277, 115
\goodbreak
\parindent 0pt
Fabbiano G., Kim D.-W., trinchieri G.: 1992 ApJS 80, 531
\goodbreak
\parindent 0pt
Fanaroff, B. L., Riley, J. M.: 1974 MNRAS 167, 31
\goodbreak
\parindent 0pt
Feretti L., Comoretto G., Giovannini G., Venturi T., Wehrle A. E.: 1993 ApJ
\goodbreak
\parindent 20pt
408, 446 ({\bf Paper III})
\goodbreak
\parindent 0pt
Garrington, S. T., Leahy, J. P., Conway, R. G., Laing, R. A.: 1988, Nature 331,
\goodbreak
\parindent 20pt
147
\goodbreak
\parindent 0pt
Garrington, S. T., Conway R. G.: 1991, MNRAS 250, 198
\goodbreak
\parindent 0pt
Ghisellini, G., Padovani, P., Celotti, A., Maraschi, L.: 1993, ApJ 407, 65
\goodbreak
\parindent 0pt
Giovannini G., Feretti L., Gregorini G., Parma P.: 1988, A\&A 199, 73
\goodbreak
\parindent 0pt
Giovannini G., Feretti L., Comoretto G.: 1990 ApJ 358, 159 ({\bf Paper I})
\goodbreak
\parindent 0pt
Grandi S. A., Osterbrock D. E.: 1978 ApJ 220, 783
\goodbreak
\parindent 0pt
Goodrich R. W., Cohen M. H.: 1992, ApJ 391, 623
\goodbreak
\parindent 0pt
Kaastra J. S., Kunieda H., Tsusaka Y., Awaki H.: 1991, A\&A 242, 27
\goodbreak
\parindent 0pt
Laing R. A.: 1988 Nature 331, 149
\goodbreak
\parindent 0pt
Laing R. A.: 1989, in Hot Spots in Extragalactic Radio Sources, ed. K.
\goodbreak
\parindent 20pt
Meisenheeimer and H. J. Roeser (Berlin: Springer), p. 27
\goodbreak
\parindent 0pt
Laing R. A.: 1993 in Astrophysical Jets, ed. Fall, M., O'Dea, C., Livio M.,
\goodbreak
\parindent 20pt
Burgarella D., p. 26
\goodbreak
\parindent 0pt
Leahy, J. P., Perley, R. A.: 1991 AJ 102, 537
\goodbreak
\parindent 0pt
Marscher A.P.: 1987 in Superluminal Radio sources, J.A. Zensus and T.J
\goodbreak
\parindent 20pt
Pearson ed., p 280.
\goodbreak
\parindent 0pt
Matthews T. A., Morgan W. M., Schmidt M.: 1964, ApJ 140, 5
\goodbreak
\parindent 0pt
Osterbrock D. E., Koski A. T., Phillips M. M.: 1976 ApJ 206, 898
\goodbreak
\parindent 0pt
Orr, M. J. L., Browne, I. W. A.: 1982 MNRAS 200, 1067
\goodbreak
\parindent 0pt
Parma P., Fanti C., Fanti R., Morganti R., de Ruiter H. R.: 1987 A\&A 181,
\goodbreak
\parindent 20pt
244
\goodbreak
\parindent 0pt
Parma P., Morganti R., Capetti A., Fanti R., de Ruiter H. R.: 1993a, A\&A
\goodbreak
\parindent 20pt
267, 31
\goodbreak
\parindent 0pt
Parma P., de Ruiter H. R., Fanti R., Laing R.: 1993b in the First Stromlo
\goodbreak
\parindent 20pt
Symposium on the Physics of Active Galaxies, in press
\goodbreak
\parindent 0pt
Pearson, T. J. 1991 BAAS, 23, 991
\goodbreak
\parindent 0pt
Pearson, T. J., Zensus J. A.: 1987, in Superluminal Radio Sources, Eds J. A.
\goodbreak
\parindent 20pt
Zensus and T. J. Pearson, Cambridge University Press, p. 1
\goodbreak
\parindent 0pt
Puschell, J. J.: 1981 AJ, 86, 16
\goodbreak
\parindent 0pt
Robertson, D. S., Ph.D. Thesis, Massachusetts Institute of Technology,
1975.
\goodbreak
\parindent 0pt
Rudy R. J., Schmidt G. D., Stockman H. S., Tokunaga A. T.: 1984 ApJ 278,
\goodbreak
\parindent 20pt
530
\goodbreak
\parindent 0pt
Spangler S. R., Pogge J. J.: 1984 AJ, 89, 342
\goodbreak
\parindent 0pt
Spangler S. R., Myers S. T., Pogge J. J.: 1984 AJ 89, 1478
\goodbreak
\parindent 0pt
Strom, R. G., Willis, A. G., Wilson, A. S.: 1978 A\&A, 68, 367,
\goodbreak
\parindent 0pt
Tadhunter, C. N., P\'{e}rez, E., Fosbury, R. A. E. 1986 MNRAS, 219, 555
\goodbreak
\parindent 0pt
Venturi T. Giovannini G., Feretti L., Comoretto G., Wehrle A. E.: 1993 ApJ
\goodbreak
\parindent 20pt
408, 81 ({\bf Paper II})
\goodbreak
\parindent 0pt
Worral D.M., Birkinshaw M.: 1994 ApJ in press
\goodbreak
\parindent 0pt
Zensus J. A.: 1993 in Jets in Extragalactic Radio Sources, Roeser H.-J and
\goodbreak
\parindent 20pt
Meisenheimer ed., p 55
\vfill
\eject
{\bf Figure Captions}
\medskip
\parindent 0pt
{\bf Fig. 1:} uv-coverage of VLBI observations. A) NGC2484; B) 3C109 C) 3C382
at 3.6 cm
\goodbreak
\parindent 0pt
{\bf Fig. 2a}: VLBI image of NGC2484 at 5.0 GHz. The HPBW is
1.50$\times$0.74 mas in PA -18$^\circ$. The noise level is 0.5 mJy/beam.
The bar on the left corner is 1 pc.
Contour values are: -1,1,2,3,4,5,7,10,15,20,30,50,100,150 mJy/beam.
\goodbreak
\parindent 0pt
{\bf Fig. 2b}: The same as Fig. 2a but super-resolved in declination;
the HPBW is 1$\times$1 mas. The bar on the left corner is 1 pc.
\goodbreak
\parindent 0pt
{\bf Fig. 3}: VLA image of 3C109 at 1.4 GHz. The HPBW is 1.5 arcsec; the noise
level is 0.2 mJy/beam.
\goodbreak
\parindent 0pt
Contour levels are:
-0.5,0.5,1,1.5,2,3,4,6,8,10,20,30,50,70,100,150,200,250,300,400,
\goodbreak
\parindent 0pt
600 mJy/beam. The map have been rotated by -51$^\circ$.
\goodbreak
\parindent 0pt
{\bf Fig. 4}: VLA image of 3C109 at 4.9 GHz with superimposed polarization
vectors. The length of
polarized lines is proportional to the polarization percentage at 5.0 GHz.
The HPBW is 1.5 arcsec; the noise level is 0.15 mJy/beam.
\goodbreak
\parindent 0pt
Contour levels are:
0.3,0.5,1,5,10,50,200 mJy/beam.
The map have been rotated by -51$^\circ$.
\goodbreak
\parindent 0pt
{\bf Fig. 5}: High resolution VLA image of 3C109 at 5.0 GHz. The HPBW is 0.55
arcsec. The noise level is 0.05 mJy/beam; contour levels are:
\goodbreak
\parindent 0pt
-0.15,0.15,0.3,0.5,0.7,1,1.5,2,2.5,3,10,30,50,100,150 mJy/beam. The
bar on the right corner is 50 kpc.
\goodbreak
\parindent 0pt
{\bf Fig. 6a}: VLBI image of 3C109 at 5.0 GHz. The HPBW is 3.0$\times$0.8 mas
in PA 165$^{\circ}$. The noise level is 0.7 mJy/beam. Contour levels are:
\goodbreak
\parindent 0pt
-1.5,1.5,3,5,7,10,20,30,50,100,150 mJy/beam. The bar on the left corner is 5
pc.
\goodbreak
\parindent 0pt
{\bf Fig. 6b}: The same as Fig. 6a but super-resolved in declination;
the HPBW is 1$\times$1 mas.
The noise level is 0.5 mJy/beam. Contour levels are:
\goodbreak
\parindent 0pt
-1.0,1.0,2,3,4,6,8,10,30,50,100,150 mJy/beam. The bar on the left corner is
5 pc.
\goodbreak
\parindent 0pt
{\bf Fig. 7}: VLBI image of 3C382 at 8.4 GHz. The HPBW is 2.0$\times$0.5 mas
in PA -6$^{\circ}$. The noise level is 0.5 mJy/beam. The bar on the left
corner is 1 pc.
\goodbreak
\parindent 0pt
Contour levels are: -1.0,1.0,1.5,2,2.5,3,4,6,8,10,15,30,50,70 mJy/beam.
\goodbreak
\parindent 0pt
{\bf Fig. 8}: Constraints on the angle $\theta$,
between the jet and the line
of sight, and the intrinsic jet velocity $\beta$.
Curves are as following: 'A' from the jet/counterjet ratio;
'B' from the core flux density (see Sect. 5); 'C' from assuming an intrinsic
maximum linear size of 1.5 Mpc; 'D' from the SSC model.
The allowed region is the undashed one.
\goodbreak
\parindent 0pt
Curve C has not been drawn for NGC2484, 3C338 and 3C382 due to the small
projected
size of these sources. The allowed region is $>$ 5$^{\circ}$ for 3C382
and $>$ 2$^{\circ}$ for the other two sources.
\bye
\magnification=1200
\baselineskip 15pt
\vsize 24.5 truecm
\hsize 15.0 truecm
\def\ltsim{\raise 2pt \hbox {$<$} \kern-1.1em \lower 4pt \hbox {$\sim$}}
\def\gtsim{\raise 2pt \hbox {$>$} \kern-1.1em \lower 4pt \hbox {$\sim$}}
\def\skuno{\vskip 15pt}
\def\skdue{\vskip 30pt}
\nopagenumbers
\parindent 0 pt
TABLE 5 - The Synchrotron-self Compton data
\skuno
\settabs 8\columns
\+ Name & ~~~~z & D$_{VLBI}$ & F$_m$ & $\nu_m$ & F$_x$
          & h$\nu_x$ & ~~$\delta$ \cr
\+      &       &  mas  & Jy      &  GHz & $\mu$Jy & kev       \cr
\+  ~~~~(1) & ~~~~(2) & (3) & (4) & (5) & (6) & (7) & ~~(8) \cr
\skuno
\+ NGC2484 & ~~0.0413 &   0.54  &  0.138  & 5.0 & \ltsim 0.28 & 1.0 & $>$0.28
\cr
\+ 3C109 &     ~~0.3066 & 0.54 & 0.175 &  5.0 & 0.32 & 2.0 & $>$0.40 \cr
\+ 3C382   &  ~~0.0578 &  0.36 & 0.116 & 8.4 & 2.2 & 2.0 & $>$0.14 \cr
\+ NGC315  & ~~0.0167 &  0.36 & 0.300 & 5.0 & 0.04 & 2.2 & $>$1.54 \cr
\+ 3C338   & ~~0.0303 &  0.54 & 0.064 & 5.0 & $<$0.21 & 2.0 & $>$0.13 \cr
\skdue
references to the X-Ray data - NGC2484: Bierman et al., 1987; 3C109, 3C382 and
3C338: Fabbiano et al., 1984; NGC315: Fabbiano et al., 1992.
\bye